\documentclass[aps,prb,preprint,groupedaddress,twocolumn,showpacs,10pt]{revtex4-1}

\usepackage{amsmath}
\usepackage{graphicx}
\usepackage{dcolumn}
\usepackage{bm}
\usepackage{hyperref}
\begin{document}


\title{Magnonic band gaps in YIG based magnonic crystals: array of grooves versus array of metallic stripes
}
\author{V. D.~Bessonov,$^{1,2}$ M.~Mruczkiewicz,$^{3}$ R. Gieniusz,$^{2}$ U. Guzowska,$^{2}$ A. Maziewski,$^{2}$ A. I. Stognij,$^{4}$ and M.~Krawczyk$^{3}$}
\email{krawczyk@amu.edu.pl}
\affiliation{
$^{1}$Faculty of Physics, University of Bia{\l}ystok, Bia{\l}ystok, Poland,\\
$^{2}$Institute of Metal Physics, Ural Division of Russian Academy of Science, Yekaterinburg, Russia,\\
$^{3}$Faculty of Physics, Adam Mickiewicz University in Poznan, Umultowska 85, Pozna\'{n}, Poland.\\
$^{4}$Scientific-Practical Materials Research Center at National Academy of Sciences of Belarus, Minsk, Belarus}

\date{\today}

\begin{abstract}
The magnonic band gaps of the two types of planar one-dimensional magnonic crystals comprised of the periodic array of the metallic stripes on yttrium iron garnet (YIG) film and YIG film with an array of grooves was analyzed experimentally and theoretically. In such periodic magnetic structures the propagating magnetostatic surface spin waves were excited and detected by microstripe transducers with vector network analyzer and by Brillouin light scattering spectroscopy. Properties of the magnonic band gaps were explained with the help of the finite element calculations. The important influence of the nonreciprocal properties of the spin wave dispersion induced by metallic stripes on the magnonic band gap width and its dependence on the external magnetic field has been shown. The usefulness of both types of the magnonic crystals for potential applications and possibility for miniaturization are discussed.  
\end{abstract}


\maketitle

\section{Introduction\label{intro}}

The spatial periodicity determines the rule of conservation of the the quasi-momentum for excitations in artificial crystals, similar to the conservation of momentum in homogeneous material.  In the frequency domain this periodicity causes the formation of the pass bands and band gaps, i.e., frequency regions in which there are no available excitation states and the wave propagation is prohibited. Magnetic structures with artificial translational symmetry are investigated to design new materials with properties that otherwise do not exist in nature, so called metamaterials. In particular, artificial ferromagnetic materials with periodicity  comparable to the wavelength of spin waves (SWs), known as magnonic crystals (MCs),\cite{Krawczyk98,Nikitov01,Puszkarski03} have recently focused attention of the physics community. The typical example of the exploitation of MCs is control of the propagation and scattering of SWs. The first experimental study of the magnetostatic SWs in ferromagnetic thin film with periodic surface was made by Sykes et al. already in 1976.\cite{Sykes1976}  Nowadays, the number of studies about MCs has surged and continues to grow at a fast pace due to interesting physics and potential new applications.\cite{Krawczyk08,Kruglyak10b,Demokritov13,Krawczyk14}

\begin{figure}[!ht]
\includegraphics[width=8cm]{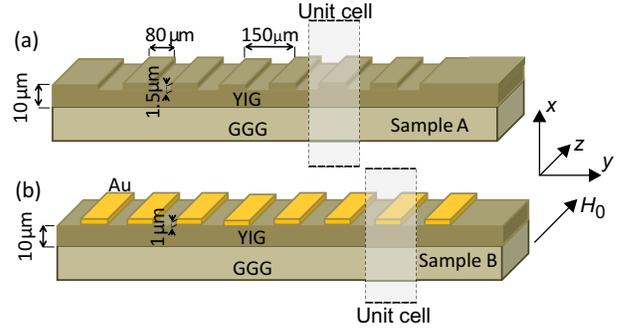}
   \caption{Geometry of the two MCs investigated in the paper. (a) Sample A, 1D MC created by the array of grooves etched in the YIG film on the GGG substrate. The thickness of the film is 10 $\mu$m, grooves width and depth is 80 $\mu$m and 1.5 $\mu$m, respectively. (b) Sample B, 1D MC formed on the basis of homogeneous YIG film of 10 $\mu$m thickness by deposition of the array of Au stripes. The stripe width is 80 $\mu$m and thickness 1 $\mu$m. The same lattice constant 150 $\mu$m is kept in both samples.}
\label{fig:Geometry}
\end{figure}

In this study we present the complementary experimental and theoretical investigation of SWs in two types of MCs having the same period in dependence on the external magnetic field amplitude. The first type is a system of periodically arranged grooves etched in the yttrium iron garnet (YIG) film, the second is an uniform YIG crystal with placed atop metallic stripes. Both types of structures have already been  studied.\cite{Gouzerh91,6028164,chumak2009scattering,chumak2009design,Filimonov2012,Kanazawa2014} The first one was proposed as a SW waveguide and experimentally tested as delay line or filter for microwave applications\cite{Sykes1976,Seshadri78,Reed85} and more recently as a basic element of the purely magnonic transistor \cite{Chumak2014} or microwave phase shifter. \cite{zhu2014magnonic} The structures of the second type were also considered as delay lines and filters\cite{Owens78,Reed85} but recently have also been proposed as room temperature magnetic field sensors.\cite{dokukin2008propagation,Inoue11:132511,Takagi2014} 

We perform comparative study of these two types of MCs magnetically saturated by the external magnetic field along grooves and metallic stripes.  For measurements we use the passive delay line with a network analyzer and Brillouin light scattering (BLS) measurements. The SW dynamic in these MCs is modeled with  finite element method (FEM) in the frequency domain. Acquired information from calculations is used to explain experimental data and to discus properties of magnonic band gap formation in these two kinds of MCs and their usefulness for recently proposed applications. 

The paper is composed as follows,  in Sec.~\ref{Sec:Exp} we briefly describe the experimental methods used in the study: measurements of the transmission of the SWs with microwave transducers and BLS measurements. In Sec.~\ref{Sec:Theory} we introduce FEM used for calculation of the magnonic band structure. In the next section, Sec.~\ref{Sec:Res}  we  discuss the results obtained in two types of MCs. The paper is ended with Sec.~\ref{Sec:Conclusions} where the summary of the paper is presented. 

\section{Experiments}\label{Sec:Exp}

The fabrication process of the artificial periodic structures with characteristic dimensions in deep nanoscale is very hard to control and so far mainly theoretical studies are available in this scale.\cite{Kim2010,Klos12}  Especially it concerns the quality of edges and the interfaces between adjacent materials which makes up MC and can significantly influence magnonic band structure.\cite{McMichael06,Klos13,Pal2014} However, in larger scale with a period starting from hundred nm the fabrication technology is already well established for thin ferromagnetic metallic films.\cite{Singh04,Adeyeye:1639} The YIG films of the thickness of tens nm have been fabricated only recently and the quality of these films increases systematically. The damping comparable to the value in thick YIG (of order smaller than in the best metallic ferromagnetic film) has already been achieved.\cite{Chang2014,Yu2014} However, the YIG thin films with patterning in nanoscale is not yet investigated. Here, we study dielectric YIG films structurized in larger scale where edge properties have minor influence on the SW dynamics. 10 $\mu$m thick YIG films were epitaxially grown on gallium gadolinium garnet (GGG) substrates in a (111) crystallographic plane and serve to fabricate one-dimensional (1D) MCs. The MCs used in our experiment had been produced in the form of the waveguide of 3.5 mm width and 50 mm  length with (i) an array of parallel grooves chemically etched [sample A, Fig. \ref{fig:Geometry}(a)] and (ii) an array of Au microstripes placed on the top of the film [sample B, Fig. \ref{fig:Geometry}(b)]. The grooves and Au microstripes were perpendicularly oriented with respect to the SW propagation direction and include nine lines of 80 $\mu$m width which are spaced 70 $\mu$m from each other, so that the period $a$ is 150 $\mu$m. 

The external magnetic field $H_0$ is applied along the grooves and microstripes in order to form conditions for propagation of the magnetostatic surface spin wave (MSSW), also called as Damon-Eshbach wave. This wave has  asymmetric distribution of the SW amplitude across the film, which depends on the direction of the magnetic field with respect to the direction of the wave vector, and this asymmetry increases with increasing wavenumber. By putting metal on the ferromagnetic film the nonreciprocal dispersion relation of MSSW is induced.\cite{Gurevich96}

\begin{figure}[!ht]
\includegraphics[width=6cm]{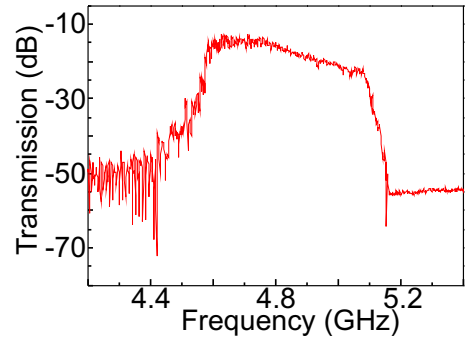}
   \caption{Transmission spectra of MSSW in the reference sample, i.e., uniform YIG film of 10 $\mu$m thickness in the external magnetic field 0.1 T.}
\label{Fig:Ref}
\end{figure}

The MSSW were excited and detected in garnet film waveguide using two 30 $\mu$m wide microstripe transducers connected with the microwave vector network analyzer (VNA), one placed in the front and another one behind the periodic structure. An external magnetic field ($\mu_0 H_0 = 0.1$ T) was  strong enough to saturate the samples. Microwave power of 1 mW used to the input transducer was sufficiently small in order to avoid any nonlinear effects. A VNA was used to measure amplitude-frequency characteristics collected for the second transducer. The transmission spectra of SWs measured in the reference sample, i.e., a thin YIG film, is shown in Fig.~\ref{Fig:Ref}. The transmission of the microwave signal above 10 dB is in the band from 4.46 to 5.14 GHz.

In BLS measurements the SWs were excited with the single 30 $\mu$m wide microwave transducer located in front of the array of Au microstripes (sample B). The SWs scattered on the line of waveguide were detected by space-resolved BLS spectroscopy in the forward scattering configuration.\cite{Demokritov2001441} The probe laser beam was scanned across the sample (in the areas between Au stripes) and the BLS intensity, which is proportional to the square of the dynamic magnetization amplitude, was recorded at various points. This technique allows for a two-dimensional (2D) mapping of the spatial distribution of the SW amplitude with step sizes of 0.02 mm.

\section{Theoretical modeling\label{Sec:Theory}}

 In our calculations we assume the stripes and grooves have infinite length (i.e., they are infinite along $z$ axis). This is reasonable assumption taking into account that length is around 23 times longer than the period of the structure.  The structures under investigation remain in a magnetically saturated state along the $z$-axis due to the static external magnetic field pointing in the same direction.

To obtain insight into the formation of the magnonic band structure and opening magnonic band gaps  the numerical calculations of the dispersion relation were performed. For SWs from the GHz frequency range, due to 10 $\mu$m thickness of YIG film and a small value of the  exchange constant in YIG, the exchange interactions can be safely neglected. In order to calculate the SW dispersion relation in magnetostatic approximation we solved the wave equation for the electric field vector ${\bf E}$:\cite{Raju06}
\begin{equation}
\nabla \times \left(\frac{1}{\hat{\mu}_{r}({\bf r})} \nabla \times {\bf E} \right) - \omega^2 \sqrt{\epsilon_{0}\mu_{0}} \left( \epsilon_{0}-\frac{i \sigma}{\omega	 \epsilon_{0}} \right) {\bf E}=0, \label{eq:main}
\end{equation}
where $\omega = 2 \pi f$, $f$ is a SW frequency, $\mu_{0}$ and $\epsilon_{0}$ denote the vacuum permeability and permittivity, respectively, and $\sigma$ is conductivity, different from zero only in sample B. To describe the dynamics of the magnetization components in the plane perpendicular to the external magnetic field, it is sufficient to solve Eq.~(\ref{eq:main}) for the $z$ component of the electric field vector ${\bf E}$ which depends solely on $x$ and $y$ coordinates: $E_{z}(x,y)$.\cite{Mruczkiewicz2014} 

The permeability tensor~$\hat{\mu}({\bf r})$ in Eq.~(\ref{eq:main}) can be obtained from Landau-Lifshitz (LL) equation.\cite{Gurevich96}
The assumption that the magnetization is in the equilibrium configuration allows us to use the linear approximation in SW calculations, which implies small deviations of the magnetization vector ${\mathbf{M}}({\mathbf{r}},t)$ from its equilibrium orientation. Thus, for the MCs saturated along $z$-axis the magnetization vector can split into the static and dynamic parts: ${\mathbf{M}}({\mathbf{r}},t)=M_{z}\hat{z}+{\mathbf{m}}({\mathbf{r}},t)$, and we can neglect all nonlinear terms with respect to dynamical components of the magnetization vector $\mathbf{m}({\mathbf{r}},t)$ in the equation of motion defined below. Since $|{\mathbf{m}}({\mathbf{r}},t)|\ll M_{z}$, we can assume also $M_{z}\approx M_{\text{S}}$, where $M_{\text{S}}$ is the saturation magnetization. We consider only monochromatic SWs propagating along the direction of periodicity,  thus we can write ${\mathbf{m}}({\mathbf{r}},t)={\mathbf{m}}(x,y) \exp(i\omega t)$. Under these assumtions the dynamics of the magnetization vector $\mathbf{m}({\mathbf{r}})$ with negligible damping is described by stationary LL equation:
\begin{eqnarray}
i \omega {\mathbf{m}}({\mathbf{r}})
=\gamma\mu_{0} \left( M_{\text{S}}\hat{z}+{\mathbf{m}}({\mathbf{r}}) \right) \times{\mathbf{H}}_{\text{eff}}({\mathbf{r}}),
\label{eq:LL}
\end{eqnarray}
where $\gamma$ is the gyromagnetic ratio (we assume $\gamma =  176$ rad GHz/T) and $\mathbf{H}_{\text{eff}}$ denotes the effective magnetic field acting on the magnetization. The effective magnetic field is in general a sum of several components, here we will consider two terms, the static external magnetic field and the dynamic magnetostatic field:
\begin{equation}
\mathbf{H}_{\text{eff}}(\mathbf{r},t)=H_{0}\hat{z}+\mathbf{h}_{\text{ms}}(\mathbf{r},t).\label{eq:Heff}
\end{equation}

The permeability tensor~$\hat{\mu}({\bf r})$ in Eq.~(\ref{eq:main}) obtained from the linearized damping-free LL Eq.~(\ref{eq:LL}) for ferromagnetic material takes following form:
\begin{equation}
\hat{\mu}_{r}=\left(
\begin{array}{ccc}
\mu^{xx} & i\mu^{xy} & 0\\
-i\mu^{yx} & \mu^{yy} & 0\\
0 & 0 & 1
\end{array}\right), \label{Eq:Permeability}
\end{equation}
where
\begin{eqnarray}
\mu^{xx} &=& \frac{\gamma \mu_{0} H_{0}(\gamma \mu_{0} H_{0}+ \gamma \mu_{0}  M_{S})-\omega^2}{(\gamma \mu_{0} H_{0})^2-\omega^2}, \\
\mu^{xy} &=&  \frac{\gamma \mu_{0}  M_{S} \omega}{(\gamma \mu_{0}H_{0} )^2-\omega^2}, \\
\mu^{yx} &=& \mu^{xy}, \;\;\;\; \mu^{yy} = \mu^{xx},
\end{eqnarray}
in non-magnetic areas permeability is an identity matrix.

Equation (\ref{eq:main}) with the permeability tensor defined in Eq.~(\ref{Eq:Permeability}) in the periodic structure  has solutions which shall fulfill Bloch theorem:
\begin{equation}
E_{z}(x,y)=E_{z}^\prime(x,y)\text{e}^{i k_{y} \cdot y },
\label{bloch}
\end{equation}
where $E_{z}^\prime(x,y)$ is a periodic function of $y$: $E_{z}^\prime(x,y) = E_{z}^\prime(x,y+a)$. $k_y$ is a wave vector component along $y$ and $a$ is a lattice constant. Due to considering SW propagation along $y$ direction only, we assume $k_y \equiv k$. Eq.~(\ref{eq:main}) together with  Eq.~(\ref{bloch}) can be written in the weak form and the eigenvalue problem can be generated, with the eigenvalues being frequencies of SWs or in the inverse eigenproblem with the wavenumbers as eigenvalues. The former eigenproblem is used to obtain magnonic band structure, the later to calculate the complex wavenumber of SW inside the magnonic band gaps. This eigenequation is supplemented with the Dirichlet boundary conditions at the borders of the computational area  placed far from the ferromagnetic film along $x$ axis (bold dashed lines in Fig.~\ref{fig:Geometry}).

In FEM the equations are solved on a discrete mesh in the two-dimensional real space [in the plane ($x$, $y$)] limited due to Bloch equation to the single unit cell (marked by the gray box in Fig.~\ref{fig:Geometry}). In this paper we use one of the realizations of FEM developed in the commercial software COMSOL Multiphysics ver.~4.2. This method has already been used in calculations of magnonic band structure in thin 1D MCs, and their results have been validated by comparing with micromagnetic simulations and experimental data.\cite{Wang09,Chi11,Mruczkiewicz13} The detailed description of FEM in its application to calculation of the SW spectra in MCs can be found in Refs.~[\onlinecite{Mruczkiewicz13,Mruczkiewicz.2013b}].

In calculations we have taken nominal values of the MC dimensions and the saturation magnetization of YIG as $M_{\text{S}} = 0.14 \times 10^6$~A/m.  The conductivity of the metal is assumed as $\sigma = 6 \times 10^7$ S/m, which is a tabular value for Au.

\section{Results and discussion \label{Sec:Res}}
\begin{figure}[!ht]
\includegraphics[width=8.5cm]{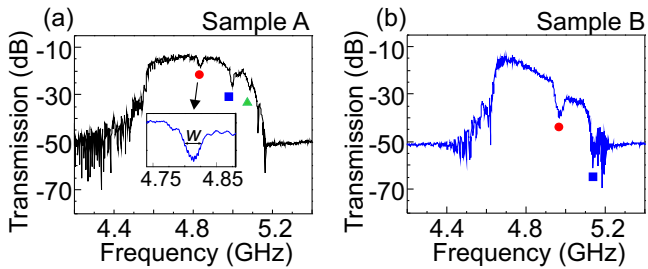}
   \caption{Transmission spectra of SWs in (a) sample A  and  (b) in sample B measured with microstripe lines in external magnetic field 0.1 T. The magnonic band gaps are marked by solid symbols: in sample A there are three gaps, in sample B there are two gaps, however the second gap is at the part of low transmission and will not be considered in the paper. In the inset of the figure (a) the enlargement of the spectra around of the first band gap is shown, the width of this gap is $w$.}
\label{Fig:Transmission}
\end{figure}

In Fig.~\ref{Fig:Transmission}(a) and (b) we present the results of the SW transmission measurements with the use of microstripe lines in the external magnetic field 0.1 T for sample A and sample B, respectively. We can see a clear evidence of three (centered at 4.81, 4.97 and 5.05 GHz) and two magnonic band gaps (at 4.88 and 5.05 GHz) in sample A and B, respectively. The transmission band in both samples is approximately the same as in the reference sample (Fig.~\ref{Fig:Ref}), however at high frequencies in sample B a large decrease of the transmission magnitude is observed. Thus, in MC with metallic stripes the second magnonic band gap [marked with blue square in Fig.~\ref{Fig:Transmission} (b)] is already at the part of the low transmission. The estimation of its position and width will be loaded with additional errors and some ambiguity, thus in further investigations we will not consider this band gap.  

\begin{figure}[!ht]
\includegraphics[width=0.4\textwidth]{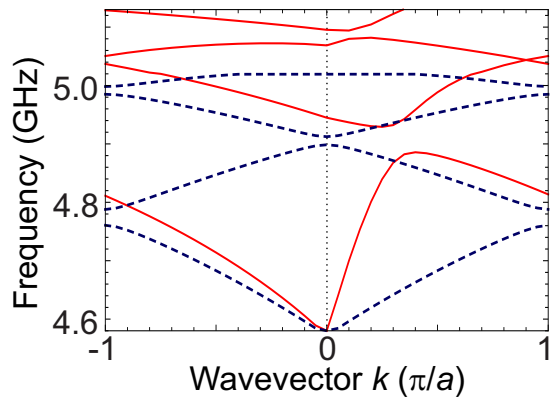}
   \caption{Magnonic band structure in the first Brillouin zone calculated for sample A (blue dashed line) and sample B (red solid line) with magnetic field $\mu_0 H_0 = 0.1$ T. Magnonic band structure is symmetric and asymmetric with respect of the Brillouin zone center (marked by vertical black dashed line) in sample A and B, respectively.}
\label{Fig:Band}
\end{figure}

 The calculated magnonic band structures are presented in Fig.~\ref{Fig:Band} with blue dashed and red solid lines for sample A and B, respectively. For MC with grooves the dispersion relation is symmetric and magnonic band gaps are opened at the Brillouin zone (BZ) border (first and third gap) and in the BZ center (the second gap). The frequencies of gaps obtained in calculations agree well with the gaps found in transmission measurements [Fig.~\ref{Fig:Transmission} (a)]. For sample B, the magnonic band structure is nonreciprocal, i.e., $f(k) \neq f(-k)$.\cite{Mruczkiewicz2014,lisenkov2014nonreciprocity} Moreover, the first band has large slope (larger than for sample A), especially in $+k$ direction has significantly increased group velocity. These effects are results of conducting properties of the Au stripes, which cause fast evanescent of the dynamic magnetic field generated by oscillating magnetization in the areas occupied by metallic stripes.   Due to this nonreciprocity in the dispersion relation the magnonic band gap opens inside the BZ and it is an indirect band gap. Also for sample B we have found good agreement between calculations and measured data.

\begin{figure}[!ht]
\includegraphics[width=0.45\textwidth]{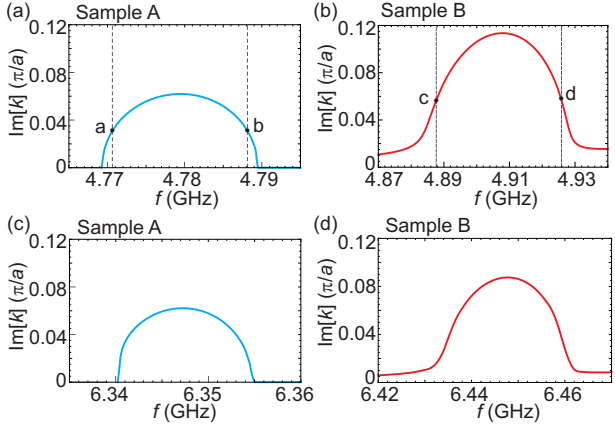}
   \caption{The imaginary part of the wavenamber around the first band gap in sample A (a), (c) and in sample B (b), (d) for the two values of the magnetic field $\mu_0 H_0 = 0.1$ T (a), (b) and 0.15 T (c), (d). The calculation were done for the inverse eigenproblem with FEM. The letters a-d indicate points where the value of Im[$k$] takes a half of its maximum. These points may indicate the borders of the band gap extracted from the transmission measurements.}
\label{Fig:Band2}
\end{figure}

In the measured data shown in Fig.~\ref{Fig:Transmission} there is visible difference between the width and depth of the first band gap in sample A and B. In order to estimate the depth of the gap from calculations we need to solve an inverse eigenproblem, i.e., to fix the frequency as a parameter and search for a complex wavenumber as an eigenvalue. In Fig.~\ref{Fig:Band2} the calculated imaginary part of the wavenumber (Im[$k$]) as a function of frequency around the first band gap is presented. In figures (a), (b) the external magnetic field was set on 0.1 T, in figures (c), (d) it was enlarged to the value 0.15 T. For sample A [Fig.~\ref{Fig:Band2}(a) and (c)] the Im[$k$] has zero value outside of the gap since the Gilbert damping is neglected in the calculations. However, for sample B [Fig.~\ref{Fig:Band2}(b) and (d)] the function Im[$k$]($f$) is nonzero outside of the gap, it is because the metal stripes induce attenuation of SWs. Outside of the gap regions in sample B the Im[$k$] increases with the frequency and this behavior is observed in the transmission spectra as decrease of the signal at large frequencies (still in the transmission band of the reference sample) in sample B [Fig.~\ref{Fig:Transmission}(b)]. 

\begin{figure}[!ht]
\includegraphics[width=7cm]{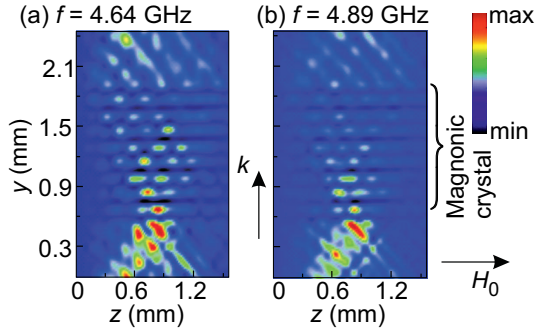}
   \caption{Maps of the SW intensity acquired with BLS from sample B at two frequencies (a) 4.64 GHz and (b) 4.89 GHz related to the transmission band and the band gap. The microstripe transducer aligned along $z$ axis used to excite SWs is located below presented area.}
\label{Fig:BLS}
\end{figure}

It is observed that the maximal value of Im[$k$] in the first band gap is significantly larger for sample B than A (0.113 and 0.062, respectively for 0.1 T magnetic field). Because an inverse of Im[$k$] describes the decaying length of SWs, it correlates with the magnonic band gap depth in the transmission measurements. Indeed this finds confirmation in the experimental data, where the minimal transmission magnitude in the band gap is  -18 dB at 4.81~GHz and -39 dB at 4.89~GHz in sample A and B, respectively. This significant suppression of the transmission of SW signal in the first magnonic band gap in sample B  is confirmed also in BLS measurements presented in Fig.~\ref{Fig:BLS}, where two excitation frequencies were set to (a) 4.64 GHz and (b) 4.89 GHz. These frequencies were chosen to visualize the SW propagation at frequencies from the band and from the band gap, respectively. In both cases the decrease of the SW amplitude with increasing the distance form the transducer is found, however in the band gap this decrease is more pronounced.  Nevertheless, some signal is still observed at the end of MC for frequencies from the band gap. We suppose that this is due to limited number of Au stripes used in the experiment and direct excitation of SWs from the transducer. 

There is also another difference between function Im[$k$]($f$) for both samples. This is an asymmetry between the bottom and top part of the gap in sample B, while in sample A the function Im[$k$]($f$) is almost symmetric with respect to the magnonic band gap center. To have some measure of this asymmetry we have calculated the derivatives  $\partial$Im[$k$]/$\partial f$ at the points where Im[$k$] is half of its maximum value, i.e., at points a-d marked in Fig.~\ref{Fig:Band2}(a) and (b). For the sample A these values are: $2.23 \times 10^{-4}$ s/m and $-2.28 \times 10^{-4}$ s/m (points a and b, respectively) and for sample B: $1.50 \times 10^{-4}$ s/m and $-1.97 \times 10^{-4}$ s/m (respectively points c and d).\footnote{The small difference of the absolute values of the derivatives at point a and b for the sample A is probably due to the asymmetry of the structure across the thickness (the grooves were etched only at one side).}  We attribute this difference in Im[$k$] between MCs to the different group velocities of SWs around the gaps, i.e., the symmetric and asymmetric dispersion curves of the first  (and second) band near the edge of the band gap for the sample A and B, respectively (Fig.~\ref{Fig:Band}). We point out that this asymmetry in Im[$k$] might appear as asymmetric slope in the transmission spectrum (Fig.~\ref{Fig:Transmission}) and it can be of some importance for applications in magnetic field sensors and magnonic transistors.\cite{Inoue11:132511,Chumak2014}

\begin{figure}[ht]
\includegraphics[width=8.5cm]{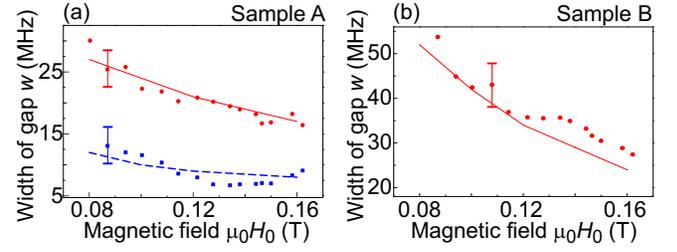}
\caption{Width of the magnonic band gap as a function of the external magnetic field (a) in sample A and (b) in sample B.  The experimental data are marked by full dots and squares, while the results of calculations are shown with solid and dashed lines for the first and second band gap, respectively. The horizontal lines at some selected values of $H_0$ show errors of the measured magnonic band gap width.}
\label{Fig:Gap_width}
\end{figure}

Finally, we study magnonic band gap widths in dependence on  amplitude of the external magnetic field. The results are presented in Fig.~\ref{Fig:Gap_width}(a) and (b) for sample A and B, respectively. In this figure there are points (full dots and squares) extracted from the transmission measurements and lines (red-solid and blue-dashed) obtained from FEM calculations (first and second band gap, respectively). Overall, we have found good agreement between theory and measurements, the calculation results are always in the range of the experimental errors marked in figures by solid vertical lines. The decrease of the band gap width with increase of the magnetic field we attribute to the decrease of the band width (i.e., decrease of the group velocity) of the MSSW.\cite{Stancil09} The steeper decrease of the band gap width is observed for the MC with metallic stripes. These dependencies find also reflection in the values of Im[$k$] shown in Fig.~\ref{Fig:Band2}, where the Im[$k$] drops down by 23\% in sample B, while for sample A the Im[$k$] remains almost the same with the increase of the magnetic field by 0.05 T. 

\begin{figure}[ht]
\includegraphics[width=4.5cm]{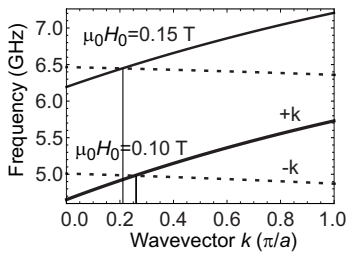}
\caption{Analytical estimation of the Bragg condition for magnonic band gap opening in 10 $\mu$m thick metalized YIG film with periodicity $a=150$  $\mu$m at two values of the external magnetic field: 0.1 T and 0.15 T. Dispersion relation of the propagating MSSW waves with maximum of the amplitude close to the metal ($k^+$) and on opposite side ($k^-$) of the film are marked with solid and dashed lines, respectively. The dispersion of $k^-$ wave is shifted by the reciprocal lattice vector $2\pi / a$ from its original position. The Bragg condition is fulfilled at the cross-section of the $k^+$ and $k^-$ lines.}
\label{Fig:estimation}
\end{figure}

This different dependencies for sample A and B are related to the larger sensitivity of the group velocity of MSSW on changes  of the magnetic field in metallized film then in unmetallized, but also to the nonreciprocal magnonic band structure and the presence of the indirect band gap in the case of sample B. The sensitivity of the group velocity  around the gap might be estimated analytically. In Fig.~\ref{Fig:estimation} the analytical dispersion relation of MSSW in 10 $\mu$m thick YIG film with metal overlayer is presented in the empty lattice model (ELM). The periodicity was taken the same as a periodicity of the samples. The crossing point between dispersions of the MSSW propagating in opposing directions, $+k$ (with maximum of the  amplitude close to the metal) and $-k$ (with amplitude on the opposite surface, in the figure this dispersion is shifted by the reciprocal lattice vector $2\pi / a$) indicates the Bragg condition, i.e., the condition for opening magnonic band gap. It means that the Bragg condition takes wavenumbers from the first and second BZ for waves propagating into positive ($+k$) and negative ($-k$) direction of the wavevector, respectively.\cite{Mruczkiewicz.2013b,Mruczkiewicz2014PRB}  The group velocities were calculated at crossing points at field values 0.1 T and 0.15 T for structures with and without metal overlayer. Based on these values, we have found that the ratio of group velocity changes with the field is almost twice higher in structure with metal layer than without this. 

Although, the change of the dispersion slope around Bragg condition is larger for $+k$ wave, the magnitude of the wavevector $|+k|$ is smaller than $|-k|$ and small change of the group velocity of $-k$ wave might also have  impact on changes in the band gap position. In our case, the increase of the magnetic field from 0.1 T to 0.15 T results also in the shift of the Bragg condition towards BZ center (see Fig.~\ref{Fig:estimation}). However, in general, the position of the Bragg condition might shift towards center or edge of the BZ with the increase of the field.  To which direction will shift the Bragg condition is determined by both group velocity change and wavenumber difference of MSSW propagating in opposite directions.  We note also, that for both samples, the width of the second band gap is less sensitive to the magnetic field amplitude than the width of the first gap.

It is also sample B which has a wider band gap then sample A in the considered magnetic field values [see Fig.~\ref{Fig:Gap_width}(a) and (b)]. It was already shown that covering bi-component MC or ferromagnetic film with lattice of grooves by a homogeneous metallic overlayer shall increase the band gap width of the MSSW due to increased group velocity of MSSW propagating along metalized surface.\cite{Sokolovskyy12,Mruczkiewicz2014PRB} However, the influence of metal with finite conductivity depends on the wavenumber (and film thickness), and disappears for large $k$.\cite{Mruczkiewicz2014} Thus, band gap width will depend on the wavenumber at which the Bragg condition is fulfilled, i.e., will depend on the lattice constant. For sufficiently large $k$ (small period) the influence of metal disappears and band gaps will not form.\cite{Mruczkiewicz2014PRB} In the homogeneous YIG film of 10 $\mu$m thickness the influence of the homogeneous Au overlayer on the dispersion relation of MSSW disappears for $k \approx 1.57 \times 10^6$ rad m$^{-1}$, i.e., for the MC with a period $a = 2$ $\mu$m at the BZ border the effect of metallization will be absent. The influence of metal will disappear also when the separation between metallic stripes and YIG will be introduced, however this can be avoided by proper fabrication technique.

  In sample A an influence of the corrugation shall preserve also for small $a$, thus it is expected that for small lattice constant the band gap in sample A will be wider than for sample B. However, we note also that the band gap width depends also on the grooves depth in sample A, thus there is an additional parameter to be taken into account. In the case of small surface perturbations (small ratio of the grove depth to the film thickness) the coupled mode theory shows that the width of the gap and the gap depth (maximal Im[$k$] in the gap) are proportional to the perturbation.\cite{Seshadri78} Nevertheless, the structure with larger groves has not been found very promissing for magnonic band gaps applications so far, because suppressed transmission in the bands due to excitations of the standing spin waves.\cite{Carter80,chumak2009scattering,chumak2009design} This can change when the very thin YIG samples will be used for MCs, then the frequency of standing exchange SW modes will moved to high frequencies. However, the fabrication regular modulation of the film thickness in deep nanoscale remain challenging task.   

\section{Conclusions}\label{Sec:Conclusions}

In summary, the SW spectrum of the two planar 1D MCs comprised of a periodic array of etched grooves in YIG film and an array of metallic stripes on homogeneous YIG film  has been fabricated and studied experimentally and theoretically. The properties of propagating SWs and magnonic band gaps were in focus of our investigations. The two different kinds of MCs elaborate the fundamental differences in the magnonic band spectrum, and also their band gap properties are different. To study of band gap widths and depths we used numerical method, which is based on FEM  in the frequency. We obtained  these values according with the measured data. We have shown that MC formed by metallic stripes posses wider magnonic band gap with larger depth than the second MC. Moreover, the fabrication of arrays of metallic stripes is much more feasible than etching of grooves in the dielectric slab. However, the influence of the metal overlayer on the band gaps of magnetostatic waves is limited to relatively small wavenumbers and this limits the miniaturizing prospective for these MCs. In contrast the MC based on the lattice of grooves does not have such limit, nevertheless its band width and depth is limiting by the excitation of the standing exchange SWs. In both types of MCs, the magnonic band gap width decreases with increasing external magnetic field, we have identified mechanisms responsible for these changes. 

The results obtained here should have impact on the applications of MCs, because we have shown the influence of different types of periodicity on the magnonic band gaps. These properties shell be especially important for magnonic devices, like magnetic field sensors\cite{Inoue11:132511} or full magnonic transistors\cite{Chumak2014} which functionality were already experimentally demonstrated. In these applications the tailoring of the magnonic band gap width, depth and their edges is crucial to make magnonic devices competitive with existing technologies. 

\begin{acknowledgments}
The research leading to these results has received funding from Polish National Science Centre project no. DEC-2-12/07/E/ST3/00538 and SYMPHONY project operated within the Foundation for Polish Science Team Programme co-financed by the EU European Regional Development Fund, OPIE 2007-2013.
\end{acknowledgments}

\bibliography{bibliography}

\end{document}